\documentclass{INTERSPEECH2023}
\usepackage{amsmath,amssymb,amsfonts}
\usepackage{algorithmic}
\usepackage{graphicx,color,graphics}
\usepackage{textcomp,comment}
\usepackage{multirow,tabularx}
\usepackage{rotating}
\usepackage[inkscapearea=page]{svg}
\usepackage{adjustbox}
\usepackage{comment,tabularx,cite}
\interspeechcameraready

\title{Severity Classification of Parkinson's Disease from Speech using \\Single Frequency Filtering-based Features}
\name{Sudarsana Reddy Kadiri, Manila Kodali, Paavo Alku}
\address{Department of Information and Communications Engineering, Aalto University, Finland.}
\email{sudarsana.kadiri@aalto.fi, manila.kodali@aalto.fi, paavo.alku@aalto.fi}

\begin{document}

\maketitle
 
\begin{abstract}
Developing objective methods for assessing the severity of Parkinson's disease (PD) is crucial for improving the diagnosis and treatment. This study proposes two sets of novel features derived from the single frequency filtering (SFF) method: (1) SFF cepstral coefficients (SFFCC) and (2) MFCCs from the SFF (MFCC-SFF) for the severity classification of PD. Prior studies have demonstrated that SFF offers greater spectro-temporal resolution compared to the short-time Fourier transform. The study uses the PC-GITA database, which includes speech of PD patients and healthy controls produced in three speaking tasks (vowels, sentences, text reading). Experiments using the SVM classifier revealed that the proposed features outperformed the conventional MFCCs in all three speaking tasks. The proposed SFFCC and MFCC-SFF features gave a relative improvement of 5.8\% \& 2.3\% for the vowel task, 7.0\% \& 1.8\% for the sentence task, and 2.4\% \& 1.1\% for the read text task, in comparison to MFCC features.

\end{abstract}

\section{Introduction}
Parkinson's disease (PD) is the second most common neurodegenerative disease after Alzheimer’s disease, and it is characterized by the progressive degeneration of dopaminergic neurons in the brain, resulting in motor and non-motor symptoms. Speech impairment is a common non-motor symptom of PD, and this impairment can significantly impact the patient's quality of life. Presence of PD and its progression is typically evaluated by a neurologist or a movement disorder specialist who assess the patient's symptoms, and overall condition over time. Regular clinical evaluations, imaging studies (such as magnetic resonance imaging or positron emission tomography scans), and other tests are utilized to monitor the progression of the disease and track the effectiveness of treatment \cite{shinde2019predictive,parkinsons_disease,rusz2018detecting}. However, these assessments are costly, laborious and prone to bias due to the neurologist's/specialist's familiarity and experience with the patients. Therefore, developing objective methods for detecting and assessing the severity of PD is crucial for improving the diagnosis, monitoring, and treatment of the disease. Speech-based severity assessment methods can be more accessible and cost-effective than traditional clinical evaluations, making them useful tools for both healthcare professionals and patients \cite{rusz2018detecting,arora2019}.

Automatic detection and severity level classification of PD from speech is facilitated by data-driven approaches based on supervised learning. This involves constructing machine learning models that are trained using speech data collected from patients and labeled by neurologists. The detection of PD (i.e., healthy $vs.$ parkinsonian) from speech has been investigated in many studies \cite{CERNAK2017196,tsanaspd1,NFL,max2009,vasquez2017convolutional,perez2019natural,erdogdu2017analyzing,parisi2018feature,moro2019forced}. More details about the various types of features and approaches used in the literature can be found in \cite{Tsanaspd2,orozco2016automatic,MEKYSKArca,cummins2018speech,hlavnivcka2017automated}. The present study focuses on speech-based severity level classification of PD (i.e., healthy $vs.$ mild $vs.$ severe).

Compared to the detection task
, much less research has been conducted in the severity level classification of PD \cite{bocklet2013automatic,arias2019predicting,arias2018unobtrusive,vasquez2019multimodal}.  In \cite{bocklet2013automatic}, the severity level (mild $vs.$ moderate $vs.$ severe) was classified using spectral, prosody, and glottal features from speech signals produced in various speaking tasks (syllable repetition tasks, read sentences, and paragraphs, as well as monologues) with the support vector machine (SVM) classifier. In \cite{arias2019predicting}, the severity level of PD was studied by grouping patients and healthy talkers into a 3-class classification problem (healthy $vs.$ mild $vs.$ affected) using perturbation, spectral, cepstral, and complexity features with deep neural network (DNN) and convolutional neural network (CNN) as classifiers. In \cite{arias2018unobtrusive}, automatic multi-class assessment was studied using a multi-class SVM following a one $vs.$ all strategy using prosody features and the monologue speaking task. In \cite{vasquez2019multimodal}, authors used onset and offset transitions from various speech sounds in continuous speech with CNNs as classifiers.

\begin{figure}[h]
\centering
\includegraphics[width=\columnwidth,height=3.3cm,trim={5cm 0.5cm, 5cm, 0cm, clip}]{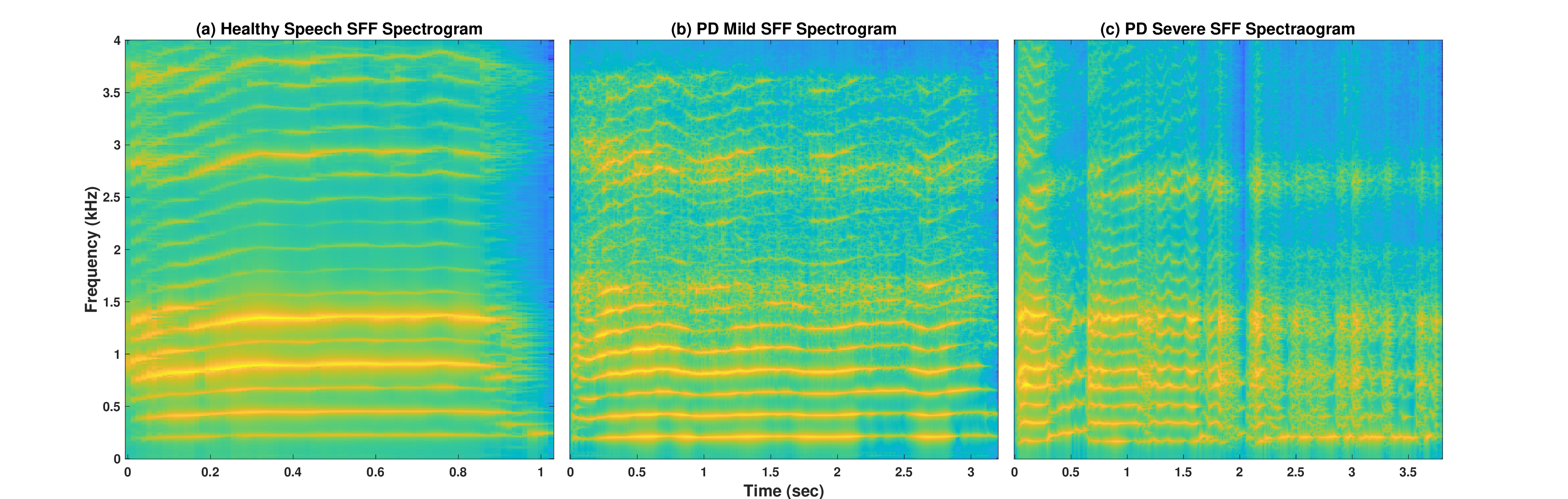}
\caption{Illustration of SFF spectrograms for  (a) healthy speech, and for PD speech of (b) mild and (c) severe severity level.}
\label{SFF_spec}
\end{figure}

\begin{figure}[ht!]
\centering
\includegraphics[width=\columnwidth,height=3.3cm,trim={5cm 0.5cm, 5cm, 0cm, clip}]{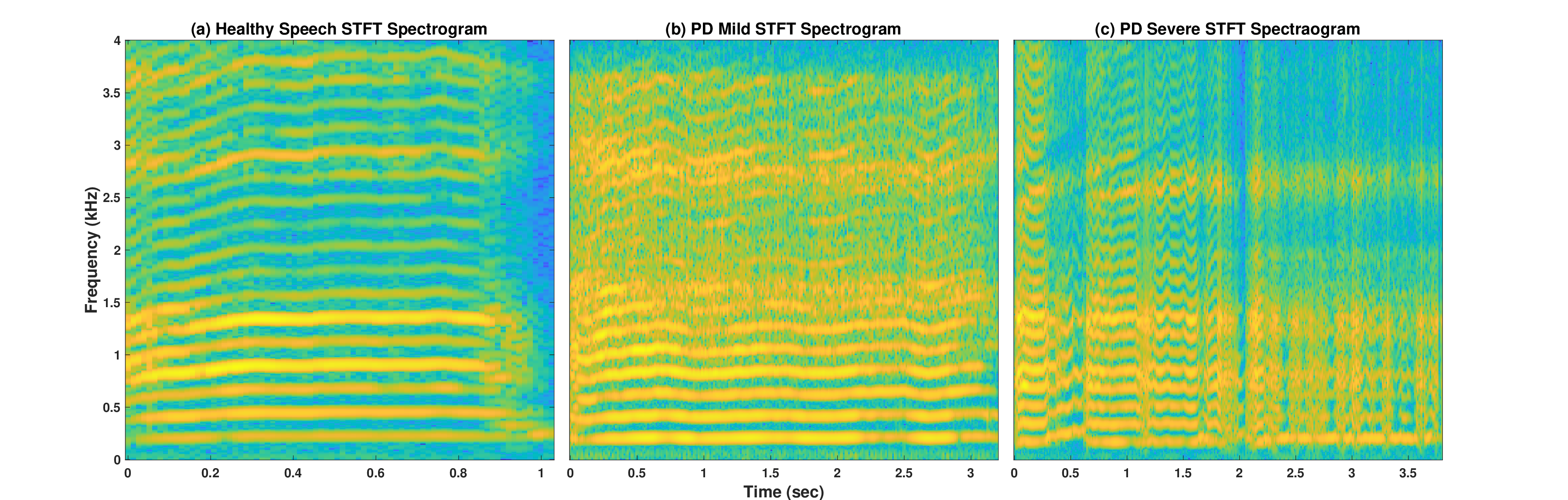}
\caption{Illustration of STFT spectrograms for  (a) healthy speech, and for PD speech of (b) mild and (c) severe severity level.}
\label{stft_spec}
\end{figure}

The results of \cite{orozco2016automatic,cepfeat,bocklet2013automatic,arias2019predicting} suggest that spectral features such as MFCCs perform better than conventional features such as phonatory/glottal and prosodic features in detection and severity classification of PD. Motivated by this, the current study investigates cepstral coefficients derived using a recently proposed signal processing method, single frequency filtering (SFF) \cite{SFF,sff_gci}. Two feature sets are derived: (1) single frequency filtering cepstral coefficients (denoted as SFFCC) and (2) MFCCs computed from the SFF spectrum (denoted as MFCC-SFF). The SFF method was shown in \cite{sff_gci,Phsing,kadiri2020parkinson} to provide higher spectral and temporal resolution for deriving speech features compared to the short-time Fourier transform (STFT), which is used in the computation of conventional MFCCs. Illustrations of spectrograms obtained with SFF and STFT are shown in Figs.~\ref{SFF_spec}~and~\ref{stft_spec}, respectively, for healthy speech and for PD speech of mild and severe levels. From the figures, it can be clearly seen that the SFF spectrogram highlights spectral information clearly (with sharper harmonics) compared to the STFT spectrogram.

The major contributions of this study are:
\begin{itemize}
\item The effectiveness of the cepstral coefficients derived from the SFF spectrum (SFFCC and MFCC-SFF) is studied for severity level classification of PD from speech.
\item A systematic comparison is conducted in severity level classification of PD between three speaking tasks (production of vowels, production of sentences as well as a text reading task).
\end{itemize}

The paper is organized as follows. Section~\ref{SFF} describes the SFF method and feature extraction from SFF method. Section~\ref{Exp} describes the experimental protocol including the database, baseline features and prior studies for comparison, classifier and evaluation metrics. Results and discussion on classification experiments are presented in Section~\ref{RnD}. A summary of the paper is given in Section~\ref{SnC}.

\section{Single Frequency Filtering (SFF)-based Features}
\label{SFF}
SFF is a time-frequency analysis technique, which provides an amplitude envelope of the speech signal as a function of time for a selected frequency \cite{SFF,sff_gci}. In this method, the speech signal $s[n]$ is first frequency-shifted (i.e., modulated) by multiplying it with an exponential function:
\begin{equation}
    \hat s[n,k]=s[n]e^{-j{2\pi\bar f_k}{n}/{f_s}},
\end{equation}
where $f_s$ is the sampling frequency, $\bar{f_k}=\frac{f_s}{2}-{f_k}$, and $f_k$ is the $k^{th}$ desired frequency. Then the frequency-shifted signal is filtered through a single-pole filter, whose root ($z=-r$) is located on the negative real axis. The transfer function of the filter is given by:
\begin{equation}
H(z)= \frac{1}{1+rz^{-1}}. 
\end{equation}
The output of the filter can be expressed as:
\begin{equation}
y[n,k] = -ry[n-1,k]+\hat s[n,k].
\vspace{-0.1cm} 
\end{equation}
The magnitude or amplitude envelope $v[n,k]$ and the phase $\psi[n,k]$ of the signal $y[k,n]$ at $k^{th}$ frequency can be written as:
\begin{equation}
 ~~~~~~~~~~ v[n,k]=\sqrt{y_{r}^2[n,k] + y_{i}^2[n,k]}, 
\end{equation} 
and 
\begin{equation}
~~~~~~~~~~ \psi[n,k]=tan^{-1}(\frac{y_{i}[n,k]}{y_{r}[n,k]}).
\end{equation} 

Here $y_r{[n,k]}$ and $y_i{[n,k]}$ are the real and imaginary parts of $y[n,k]$, respectively. The amplitude envelope of the signal can be computed for several frequencies at intervals of $\Delta{f}$ by defining ${f_k}$ as follows:
\begin{equation}
 f_{k}=k\Delta{f},     ~~~~~~~~~k=1,2,\ldots, K, 
\end{equation}
where $K=\frac{({f_s}/2)}{\Delta{f}}.$  
The SFF magnitude spectrum can be obtained for each instant of time from $v[n,k]$.

\subsection{Extraction of the SFFCC features}
\label{sec:SFFCCs}
\begin{figure}[ht]
\centering
\includegraphics[width=0.5\textwidth,height=1.5cm]{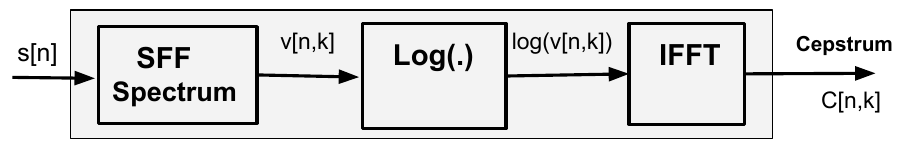}
\vspace{-0.5cm}
\caption{\label{SFFCCs}Block diagram of the extraction of single frequency filtering cepstral coefficients (SFFCC).}
\end{figure}

SFFCCs are extracted by computing the cepstrum ($C[n,k]$). Cepstrum is derived from the SFF spectrum ($v[n,k]$) as follows:
	     \begin{equation}
	    C[n,k]= {\rm IFFT}(\log (v[n,k])), 
	      \vspace{-0.1cm}
	     \end{equation}	        
where IFFT is the inverse Fourier transform. From cepstrum $C[n,k]$, the first 13 cepstral coefficients (including $0^{th}$) are considered and they are referred to as SFFCC. Apart from the static coefficients, the delta ($\Delta$) and double-delta ($\Delta\Delta$) coefficients are also computed, which results in a 39-dimensional feature vector. A schematic block diagram describing the steps involved in the extraction of SFFCC is shown in Fig.~\ref{SFFCCs}.

\subsection{Extraction of the MFCC-SFF features}
\begin{figure}[ht]
\centering
\includegraphics[width=0.5\textwidth,height=1.7cm]{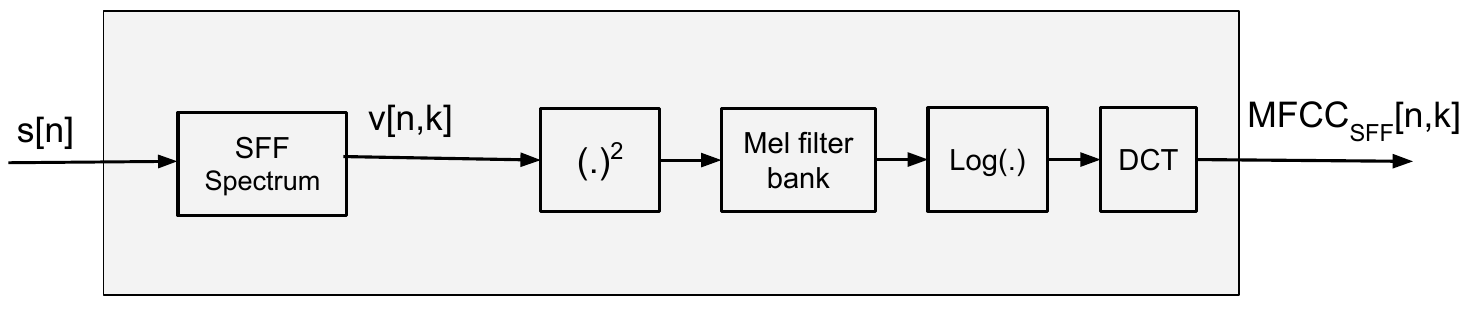}
\caption{\label{MFCC-SFF}Block diagram of the extraction of MFCCs from the SFF spectrum (MFCC-SFF) \cite{kethireddy2022deep}.}
\end{figure}

A schematic block diagram describing the steps involved in the extraction of MFCCs from the SFF spectrum is shown in Fig.~\ref{MFCC-SFF}. The MFCC extraction consists of mel filter$-$bank analysis on the SFF spectrum followed by logarithm and discrete cosine transform (DCT) operations. This is expressed as follows:
\begin{equation}
MFCC_{SFF}[n,k]= DCT(\log(Mel({ S_{SFF}[n,k]}^2))),
\end{equation}	
where $MFCC_{SFF}[n,k]$ denotes the mel-cepstrum. The resulting cepstral coefficients are referred as MFCC$-$SFF, and they represent compactly the spectral characteristics of speech. From the mel-cepstrum, the first 13 cepstral coefficients (including the $0^{th}$) are considered. Apart from the static coefficients, the $\Delta$ and $\Delta\Delta$ coefficients are also computed, which results in a 39-dimensional feature vector.

\section{Experimental Protocol}
\label{Exp}
This section describes the speech database used in the experiments, the reference features that were selected for comparison and the classifier.

\subsection{Database} 
\begin{figure}[h]
\centering
\begin{tabular}{cc}
\includegraphics[width=0.5\linewidth,height=4cm]{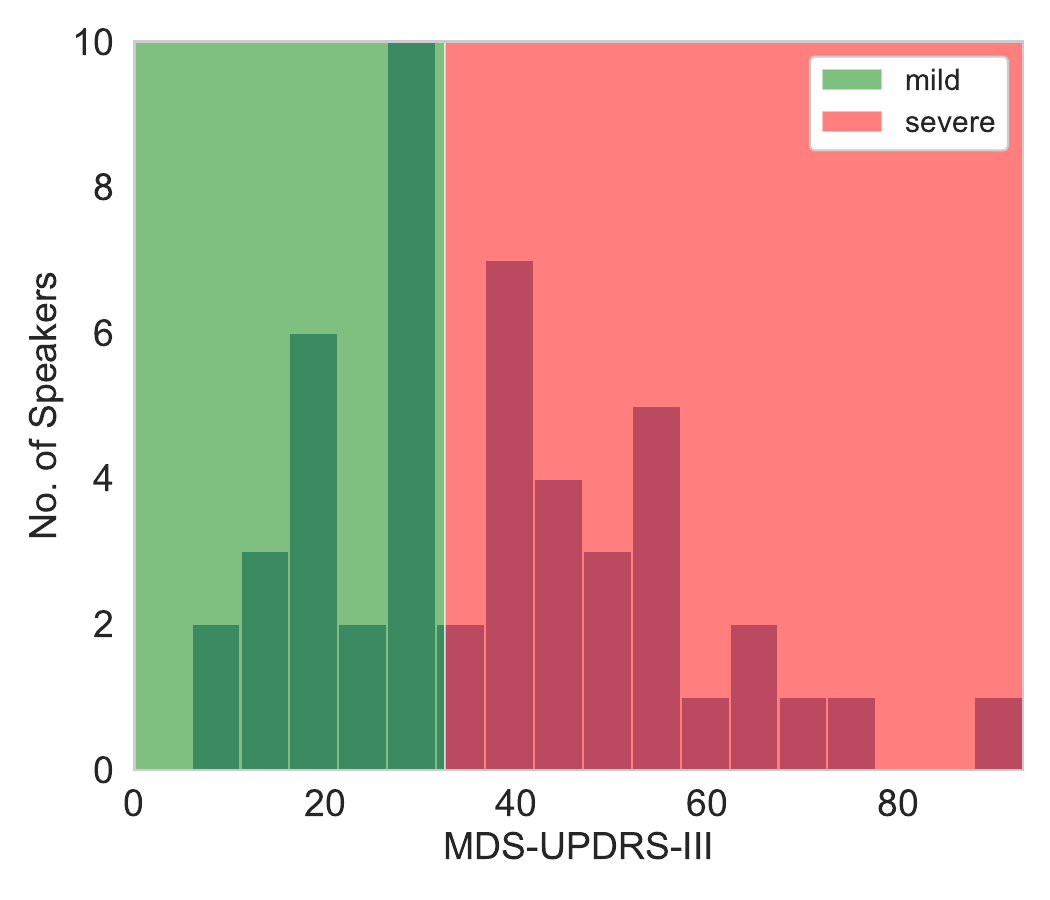}&
\includegraphics[width=0.5\linewidth,height=4cm]{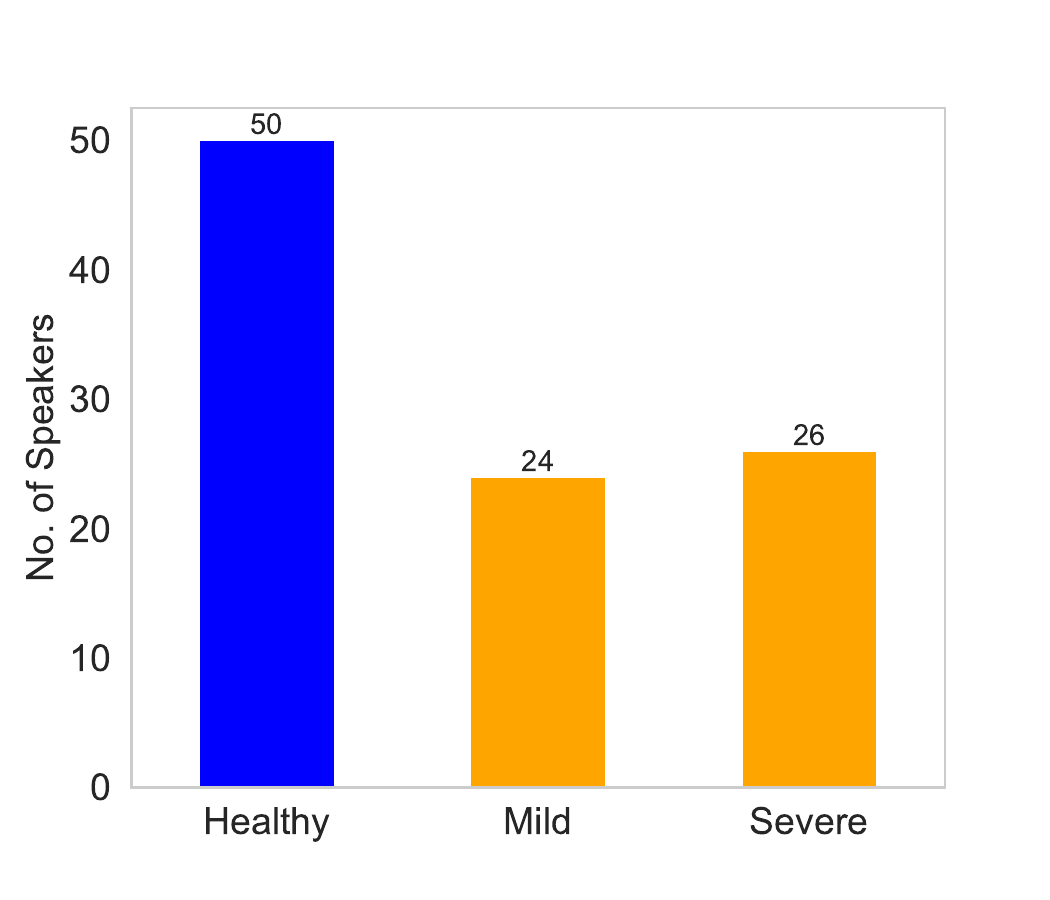}\\
\end{tabular}
\vspace{-0.4cm}
\caption{Distribution of the MDS-UPDRS III (left) and the number of speakers available in the database for each severity class (right).}
\label{db}
\end{figure}

This study uses publicly available PC-GITA, a repository of Spanish parkinsonian speech \cite{spanishPDcorpus}. The database is balanced with respect to gender and age, and it consists of speech recordings from 50 PD patients (25 female and 25 male) and 50 healthy control speakers (HCs) (25 female and 25 male). For the male PD patients, the age range is between 33 and 77 years (mean 62.3 years), for the female PD patients, the age range is between 44 and 75 years (mean 60.2 years). For the male HCs, the age range is between 31 and 86 years (mean 61.3 years), and for the female HCs, the age range is between 43 and 76 years (mean 60.8 years). The database consists of different speaking tasks including productions of vowels, sentences, diadochokinetic words, as well as text reading and a monologue task. In this study, we considered the vowel task, the sentence task and the read text task. The vowel task includes three repetitions of five Spanish vowels. The sentence task includes productions of six different sentences. The read text task consists of a dialog between a doctor and a speaker, and this task is phonetically balanced (36 words). All the patients were diagonsed by neurologists, and their disease severity was labeled according to the modified H \& Y scale and the MDS-UPDRS-III. The MDS-UPDRS-III scale is based on the motor examination and it consists of 33 items, where each of the item ranges from 0 (normal) to 4 (severe). The database was partitioned into three severity classes (healthy, mild, and severe) according to the MDS-UPDRS-III scores. Figure~\ref{db} shows the distribution of the MDS-UPDRS-III and the number of speakers per each class. This study uses the balanced number of speakers for each class. Further details of the database can be found in \cite{spanishPDcorpus,rusz2013imprecise}.

\vspace{-0.3cm}
\subsection{Baseline features and reference classifiers of a prior study}
\label{ref_feats}
The most popular speech features, namely MFCCs \cite{LPfeatpat,henriquez2009characterization,orozco2016automatic,bocklet2013automatic}, are used as the baseline features. The first 13 MFCCs (including $0^{th}$) are extracted using a Hamming window size of $30$ ms and a shift of $10$ ms. Apart from the above static coefficients, the $\Delta$ and $\Delta\Delta$ coefficients are also computed resulting in a 39-dimensional feature vector. In addition to the baseline MFCCs, we compared the results of the current study also with the results reported in  \cite{arias2019predicting}, where a protocol similar to this study was utilized using only the vowel task. In \cite{arias2019predicting}, perturbation, spectral, cepstral and complexity features were used together with a DNN, and the modulation spectrum was used with a CNN and multi-modal architectures.

\subsection{Parameters used for the SFF-based features}
For the SFF spectrum estimation, $r=0.99$ and $\Delta{f}=31.25$ Hz (resulting in 512 amplitude envelopes) are used. SFFCC and MFCC-SFF are extracted with an interval of $10$ ms rather than considering every time instant. The number of mel-filters used is 80, and the first 39 cepstral coefficients are extracted for both of the features. Frame-wise features (for baseline and proposed) are merged into a 1-dimensional feature vector by taking the mean.

\subsection{Classifier and evaluation metrics}
\label{eval_metrics}
Severity classification experiments were carried out using the SVM classifier (linear kernel in the one-vs-one architecture, with c and gamma parameters in the range of $10^{-4}$ to $10^{4}$), which is known to be an effective classifier when the amount of training data is limited. The experiments were conducted with leave-one-speaker-out (LOSO) cross-validation, where one speaker's data was considered as a test set and the remaining speakers' data was used for training the classifier. In each fold, the evaluation metrics were saved, and the procedure was repeated for all the speakers, and finally the evaluation metrics were averaged. 

The metrics chosen are the balanced classification accuracy (also known as unweighted average recall (UAR)), class-wise precision, class-wise recall, and class-wise F1 score. We also used confusion matrices for assessing the performance of the classification systems.

\section{Results and Discussion}
\label{RnD}

\begin{table*}[t]
\centering
\caption{\label{Result_ST}Results of Parkinson's disease severity classification for the reference MFCC features, the proposed SFFCC features and the MFCC-SFF features in three speaking tasks (vowels, sentences and read text). Precision, recall and F1 score are given for individual classes (0 for healthy, 1 for mild, and 2 for severe).}
\vspace{-0.4cm}
\resizebox{\textwidth}{2.8cm}{
\begin{tabular}{|lllllllllll|}
\hline
\multicolumn{11}{|c|}{{\bf Vowel task}}\\ \hline \hline
\multicolumn{1}{|l|}{Feature}  & \multicolumn{1}{l|}{Accuracy}   & \multicolumn{1}{l|}{Precision-0} & \multicolumn{1}{l|}{Recall-0} & \multicolumn{1}{l|}{F1 score-0} & \multicolumn{1}{l|}{Precision-1} & \multicolumn{1}{l|}{Recall-1} & \multicolumn{1}{l|}{F1 score-1} & \multicolumn{1}{l|}{Precision-2} & \multicolumn{1}{l|}{Recall-2} & F1 score-2 \\ \hline
\multicolumn{1}{|l|}{MFCC}     & \multicolumn{1}{l|}{48.7 ± 7}   & \multicolumn{1}{l|}{0.53}        & \multicolumn{1}{l|}{0.57}     & \multicolumn{1}{l|}{0.55}       & \multicolumn{1}{l|}{0.40}        & \multicolumn{1}{l|}{0.38}     & \multicolumn{1}{l|}{0.39}       & \multicolumn{1}{l|}{0.45}        & \multicolumn{1}{l|}{0.45}     & 0.45       \\ \hline
\multicolumn{1}{|l|}{SFFCC}    & \multicolumn{1}{l|}{\bf 51.5 ± 4}   & \multicolumn{1}{l|}{0.50}        & \multicolumn{1}{l|}{0.56}     & \multicolumn{1}{l|}{0.52}       & \multicolumn{1}{l|}{0.43}        & \multicolumn{1}{l|}{0.40}     & \multicolumn{1}{l|}{0.42}       & \multicolumn{1}{l|}{0.49}        & \multicolumn{1}{l|}{0.46}     & 0.47       \\ \hline
\multicolumn{1}{|l|}{MFCC-SFF} & \multicolumn{1}{l|}{49.8 ± 5}   & \multicolumn{1}{l|}{0.49}        & \multicolumn{1}{l|}{0.54}     & \multicolumn{1}{l|}{0.52}       & \multicolumn{1}{l|}{0.42}        & \multicolumn{1}{l|}{0.42}     & \multicolumn{1}{l|}{0.42}       & \multicolumn{1}{l|}{0.45}        & \multicolumn{1}{l|}{0.42}     & 0.44       \\ \hline \hline
\multicolumn{11}{|c|}{\bf Sentence task} \\ \hline \hline
\multicolumn{1}{|l|}{MFCC}     & \multicolumn{1}{l|}{54.5 ± 8}   & \multicolumn{1}{l|}{0.53}        & \multicolumn{1}{l|}{0.61}     & \multicolumn{1}{l|}{0.57}       & \multicolumn{1}{l|}{0.52}        & \multicolumn{1}{l|}{0.45}     & \multicolumn{1}{l|}{0.48}       & \multicolumn{1}{l|}{0.63}        & \multicolumn{1}{l|}{0.62}     & 0.62       \\ \hline
\multicolumn{1}{|l|}{SFFCC}    & \multicolumn{1}{l|}{\bf 58.3 ± 7.8} & \multicolumn{1}{l|}{0.60}        & \multicolumn{1}{l|}{0.63}     & \multicolumn{1}{l|}{0.62}       & \multicolumn{1}{l|}{0.53}        & \multicolumn{1}{l|}{0.51}     & \multicolumn{1}{l|}{0.52}       & \multicolumn{1}{l|}{0.62}        & \multicolumn{1}{l|}{0.61}     & 0.62       \\ \hline
\multicolumn{1}{|l|}{MFCC-SFF} & \multicolumn{1}{l|}{55.5 ± 8}   & \multicolumn{1}{l|}{0.58}        & \multicolumn{1}{l|}{0.60}     & \multicolumn{1}{l|}{0.59}       & \multicolumn{1}{l|}{0.54}        & \multicolumn{1}{l|}{0.53}     & \multicolumn{1}{l|}{0.54}       & \multicolumn{1}{l|}{0.62}        & \multicolumn{1}{l|}{0.61}     & 0.61       \\ \hline \hline
\multicolumn{11}{|c|}{\bf Read text task} \\ \hline \hline
\multicolumn{1}{|l|}{MFCC}     & \multicolumn{1}{l|}{61.5 ± 10}  & \multicolumn{1}{l|}{0.59}        & \multicolumn{1}{l|}{0.79}     & \multicolumn{1}{l|}{0.68}       & \multicolumn{1}{l|}{0.45}        & \multicolumn{1}{l|}{0.42}     & \multicolumn{1}{l|}{0.43}       & \multicolumn{1}{l|}{0.50}        & \multicolumn{1}{l|}{0.38}     & 0.43       \\ \hline
\multicolumn{1}{|l|}{SFFCC}    & \multicolumn{1}{l|}{\bf 63.0 ± 12}  & \multicolumn{1}{l|}{0.55}        & \multicolumn{1}{l|}{0.75}     & \multicolumn{1}{l|}{0.63}       & \multicolumn{1}{l|}{0.44}        & \multicolumn{1}{l|}{0.33}     & \multicolumn{1}{l|}{0.38}       & \multicolumn{1}{l|}{0.62}        & \multicolumn{1}{l|}{0.54}     & 0.58       \\ \hline
\multicolumn{1}{|l|}{MFCC-SFF} & \multicolumn{1}{l|}{62.2 ± 13}  & \multicolumn{1}{l|}{0.58}        & \multicolumn{1}{l|}{0.62}     & \multicolumn{1}{l|}{0.60}       & \multicolumn{1}{l|}{0.46}        & \multicolumn{1}{l|}{0.46}     & \multicolumn{1}{l|}{0.46}       & \multicolumn{1}{l|}{0.50}        & \multicolumn{1}{l|}{0.46}     & 0.48       \\ \hline
\end{tabular}}
\end{table*}

\begin{figure*}[ht]
\centering
\begin{tabular}{ccc}
\bf{~{MFCC}} & \bf{~~~~~{SFFCC}} & \bf{~~~~~~~~{MFCC-SFF}}\\
\rotatebox{90}{\hspace{1.1cm}\large Actual class\vspace{-.5cm}}
\includegraphics[width=0.31\linewidth,height=4.5cm,trim=1cm 1cm 0cm 0cm, clip]{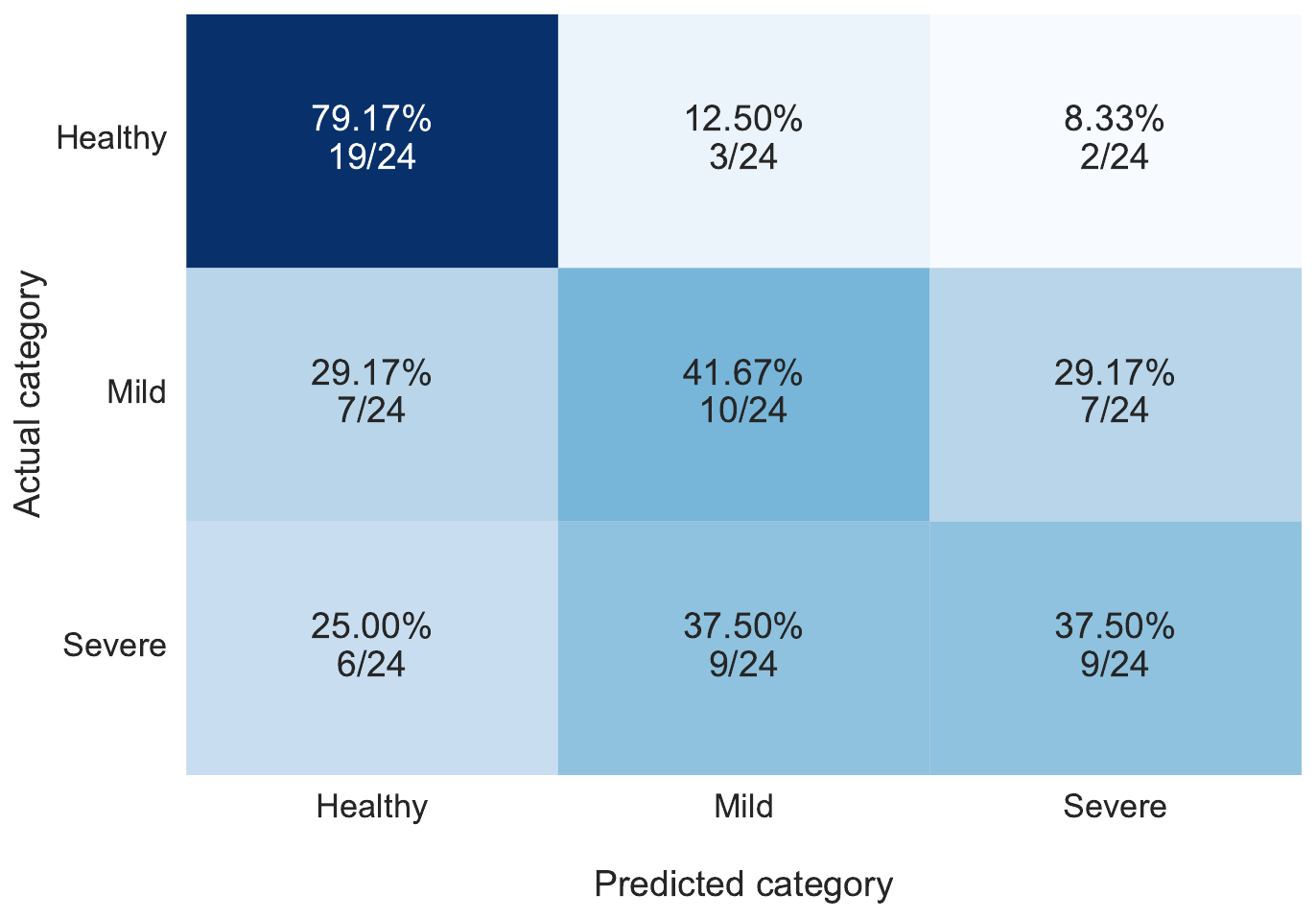}&
\includegraphics[width=0.31\linewidth,height=4.5cm,trim=1cm 1cm 0cm 0cm, clip]{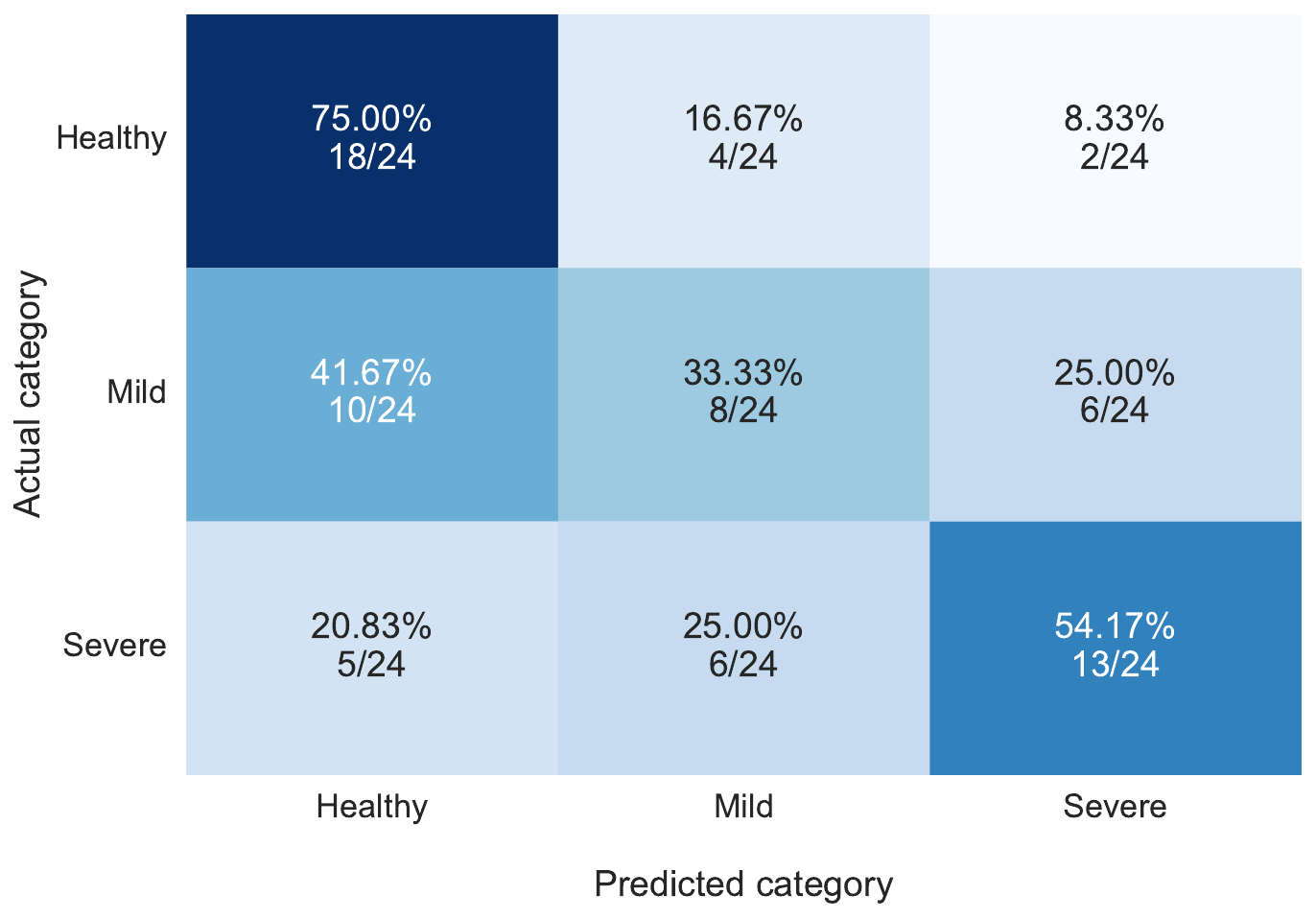}&
\includegraphics[width=0.31\linewidth,height=4.5cm,trim=1cm 1cm 0cm 0cm, clip]{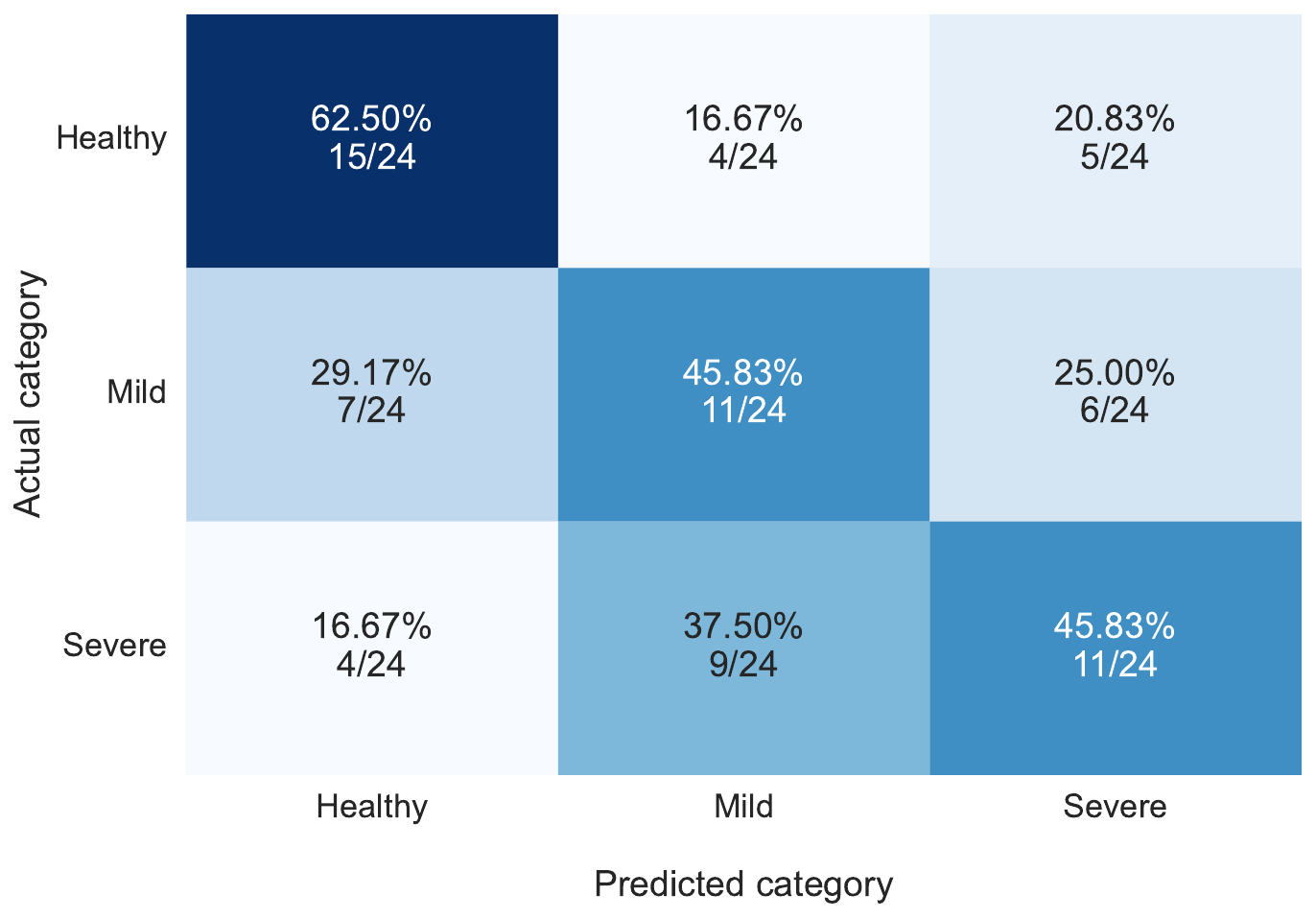}\\
\end{tabular}
{\large Predicted class} 
\vspace{-0.1cm}
\caption{\label{confmat}Confusion matrices for the MFCC (left), SFFCC (middle) and MFCC-SFF (right) features in the read text task. Rows correspond to the actual Parkinson's disease severity classes and columns correspond to the predicted classes.}
\vspace{-0.1cm}
\end{figure*}

\begin{table}[ht]
\centering
\caption{\label{Result_ST1}Classification accuracy in the vowel task based on the features proposed in the current study and based on classifiers studied in a previous study. 
}
\vspace{-0.3cm}
\begin{tabular}{lllll|}
\cline{3-5}
                                     & \multicolumn{1}{l|}{}         & \multicolumn{3}{l|}{Class-wise accuracies}                         \\ \hline
\multicolumn{1}{|l|}{}               & \multicolumn{1}{l|}{Accuracy} & \multicolumn{1}{l|}{Healthy} & \multicolumn{1}{l|}{Mild}  & Severe \\ \hline
\multicolumn{5}{|c|}{\textbf{Current study}}                                                                                             \\ \hline
\multicolumn{1}{|l|}{SFFCC (SVM)}    & \multicolumn{1}{l|}{51.5 $\pm$ 4}   & \multicolumn{1}{l|}{55.9}    & \multicolumn{1}{l|}{40.4} & 45.6   \\ \hline
\multicolumn{1}{|l|}{MFCC-SFF (SVM)} & \multicolumn{1}{l|}{49.8 $\pm$ 5} & \multicolumn{1}{l|}{54.2}    & \multicolumn{1}{l|}{40.9}  & 42.5   \\ \hline
\multicolumn{5}{|c|}{\textbf{Previous study \cite{arias2019predicting}}}                                                                           \\ \hline
\multicolumn{1}{|l|}{DNN }            & \multicolumn{1}{l|}{50.0 $\pm$ 9}   & \multicolumn{1}{l|}{30.9}    & \multicolumn{1}{l|}{43.6}  & 70.0     \\ \hline
\multicolumn{1}{|l|}{CNN }            & \multicolumn{1}{l|}{41.0 $\pm$ 5}   & \multicolumn{1}{l|}{68.0}      & \multicolumn{1}{l|}{21.8}  & 33.3   \\ \hline
\multicolumn{1}{|l|}{Multi-modal}    & \multicolumn{1}{l|}{52.0 $\pm$ 7}   & \multicolumn{1}{l|}{63.9}    & \multicolumn{1}{l|}{40.0}    & 46.7   \\ \hline
\end{tabular}
\end{table}

This section reports the results of the experiments for the three speaking tasks (vowels, sentences and read text) by first describing the classification accuracies obtained and then describing the confusion matrices. Table~\ref{Result_ST} shows the results in terms of the mean and standard deviation of accuracy along with class-wise precision, recall and F1-score. From the results in the table, it can be observed that the performance of both of the proposed features (SFFCC and MFCC-SFF) is clearly better than the baseline MFCCs features for all the three speaking tasks. From the speaking tasks, the performance of the read text task is better that of the sentence and vowel tasks. Between the sentence and vowel tasks, the former task gave better results. Taken together, the order of preference for the speaking tasks is: \\

{$Read~Text$ $>$ $Sentences$ $>$ $Vowels$.} \\

This trend was expected as the read text task included richer articulation information (due to phonetically balanced sounds) which is helpful for severity classification. From the metrics based on precision, recall and F1 score, it is evident that the MFCC features are more biased towards the healthy class (specifically in the vowel and read text tasks) compared to the proposed features. Overall, the proposed SFFCC and MFCC-SFF features gave an absolute improvement of 2.8\% and 1.1\% (relative improvement of 5.8\% and 2.3\%) for the vowel task, 3.8\% and 1.0\% (relative improvement of 7.0\% and 1.8\%) for the sentence task, and 1.5\% and 0.7\% (relative improvement of 2.4\% and 1.1\%) for the read text task, in comparison to the baseline MFCC features. Between the two proposed features, SFFCC performed better than MFCC-SFF.

Figure~\ref{confmat} shows the confusion matrices for the MFCC (left), SFFCC (middle) and MFCC-SFF (right) features in the read text task. This figure shows clearly that all the features are biased towards healthy class. Among the features, the proposed features are, however, less biased towards health class compared to the MFCC features. This is also in conformity with the performances reported in Table~\ref{Result_ST}.

Table~\ref{Result_ST1} reports the performance of the proposed SFFCC and MFCC-SFF features for the vowel task by showing the results of the previous reference study \cite{arias2019predicting}. It can be seen that the performance of the proposed features is better than in the reference study (except for the multi-modal system). In addition, we would like to note that the standard deviations in accuracy are lower for the proposed features in comparison to the prior study. The results of the experiments indicate that the proposed features effectively capture speech articulation variations which reflect changes in the disease severity.

\vspace{-0.3cm}
\section{Summary}
\label{SnC}
\vspace{-0.2cm}
In this study, we proposed two sets of features derived from the SFF spectrum (SFFCC and MFCC-SFF) for the severity classification of PD from speech. The classification experiments were carried out with three speaking tasks (sustained vowels, sentences and read text) of the well-known PC-GITA database. The experiments with the SVM classifier revealed that the proposed SFFCC and MFCC-SFF features outperformed the conventional MFCCs features in all three speaking tasks. Among the speaking tasks, the classification performance was highest for the text reading task and lowest for the vowel production task. This trend is due to the larger diversity of articulation information in speech signals produced in the text reading task (due to phonetically balanced sounds), which is helpful for severity classification. 
\vspace{-0.3cm}
\section{Acknowledgements}
This study was funded by the Academy of Finland (project no. 330139).



\end{document}